\documentstyle [11pt,psfig]{article}
\begin{document}
\begin{center}
{\bf Evolution of Baryon Rich Quark-Gluon Plasma and radiation of Single Photons}
\\
{S. V. S. Sastry, D. Dutta, A. K. Mohanty  and  D. K. Srivastava$^*$}\\
{ Nuclear Physics Division, Bhabha Atomic Research Centre,
Trombay, Mumbai 400 085, India\\
$^*$ VECC, 1/AF Bidhan Nagar, Kolkata 700 064, India}\\
$^*$ Duke University, Dept. of Physics, Box 90305, Durham, NC 27708-0305.
\vspace {-.4cm}
\end{center}

\par
\vspace {1.5cm}
\centerline{Abstract}
\par

\noindent
The (3+1) dimensional expansion of the quark gluon plasma (QGP) produced at
finite  baryon  density  has been studied using relativistic hydrodynamical
approach. The pressure functional of the equation of state (EoS)  has  been
determined for the interacting nuclear matter with mesons exchange. The EoS
has  been used to solve hydrodynamical equations using RHLLE algorithm. The
space time expansion of the plasma has been studied for the  cases  of  SPS
energy and RHIC energy both at finite baryon density and for a hypothetical
case  of  SPS  energy  at  zero baryon density. The space-time evolution is
slowed and the life times of QGP and mixed  phases  are  shortened  in  the
presence  of  finite baryon density. The space time integrated total photon
yields have been estimated by convoluting the static  emission  rates  with
the  space  time  expansion of the plasma. It has been shown that the total
photon yield at zero rapidity is not significantly affected by  the  baryon
density  for  SPS  energy.  The  total  photon  yield  is unaffected by the
Landau-Pomeranchuk-Migdal effect at SPS energy at zero baryon  density,  as
the  quark matter contribution to photon yield is less compared to hadronic
matter.

\par

\vspace {2.cm}

\noindent
Electromagnetic  processes  such  as  photons  and dileptons production are
important to identify the quark gluon plasma (QGP) expected to be formed in
the ultra relativistic heavy ion collisions at SPS, RHIC and  LHC  energies
\cite{peitz,alam1,dks1,dks2}.   Consequent   to  the  availability  of  the
experimental data reported by  WA98  collaboration  \cite{wa98}  of  single
photons  from  Pb+Pb  collisions  at  CERN  SPS,  these studies gained much
interest in recent times. The experimental spectra however include  photons
coming  from  various  sources  like  decays,  prompt  and thermal photons.
Photons are produced at each stage from an expanding volume of plasma being
both in QGP and hadron phases. The yields of these electromagnetic  signals
are  quite  sensitive  to  the  space time evolution of the QGP. Assuming a
first order phase transition, the plasma will expand and evolve  through  a
QGP, mixed and hadron phases governed by the appropriate equation of state.
Thermal  photons  are  emitted  at a later stage which depends on the space
time evolution scenario considered  consequently,  the  expansion  dynamics
determines   the   observed   spectra.   Initially,   the   plasma  expands
longitudinally, however at high energy densities and when plasma life  time
is  large,  the transverse expansion effects become significant and can not
be ignored \cite{alam2}.  Further,  the  experimental  data  indicates  the
presence  of  finite baryon density at freezeout temperature. The evolution
becomes more complicated if the plasma is unsaturated or formed  at  finite
baryon  density.  These  physical  conditions of the plasma affect both the
basic photon production rate and  also  the  total  space  time  integrated
photon yields. In the present work we incorporate the finite baryon density
effects in both the production rates and the plasma expansion dynamics.

\noindent
The photon production rates from QGP phase have been investigated in detail
in \cite{kapusta}-\cite{dutta1}. These studies include the production rates
at  effective  one  loop  and  two  loop  levels.  Further,  multiple  soft
scatterings of  the  fermion  during  photon  emission  reduce  the  photon
coherence  lengths,  known  as  Landau-Pomeranchuk-Migdal (LPM) effect, and
suppress the production  rates  \cite{auren2,arnold1,arnold2}.  The  photon
rates  from  the  hadron phase were estimated considering the comprehensive
set of processes in  \cite{kapusta,sarkar},  and  the  prompt  decays  were
estimated  in  \cite{wwang,gall}.  Several groups have already reported the
space-time integrated photon yield  calculations  for  SPS,  RHIC  and  LHC
energies   considering  (i)  various  sets  of  equations  of  state,  (ii)
considering one or three dimensional space time expansion of plasma,  (iii)
with     or     without     boost     invariance,     for    details    see
\cite{peitz,alam1,dks1,dks2,alam2,huovinen,alam3}.

\par
\noindent
In the following, the expansion dynamics of the plasma is determined by the
EoS   and  the  relativistic  hydrodynamical  equations.  The  conservation
equations for the energy ($E$), momentum ($M$) and net baryon number  ($R$)
densities are given by,

\begin{equation}
\partial _{\mu}T^{\mu\nu} = 0 ~~~~\mbox{and,}~~~~\partial _{\mu}N^\mu = 0
\end{equation}
\begin{equation}
 E = T^{00} = (\epsilon + p)\gamma^2 -p
\end{equation}
\begin{equation}
{\bf M} =T^{0i} = (\epsilon + p)\gamma^2{\bf v}
\end{equation}
\begin{equation}
R = N^0 = nu^0 = n\gamma
\end{equation}

\noindent
Here,  $u^\mu=\gamma(1,{\bf  v}  )$  is  the  fluid velocity. In the above,
$\epsilon,p,n$ are the energy, pressure and net baryon number densities  in
the local rest frame of the fluid. The conservation equations for $ E,M,R $
of Eqs. (2-4) for cylindrical geometry with the longitudinal (z-axis) boost
invariance  and at z=0 are as discussed in \cite{rhlle1}. The shock just at
the QGP formation time is represented by the Riemann initial value problem.
The full solution to the Riemann problem is  approximated  by  the  Godunov
scheme consisting of a rarefaction wave and the contact discontinuity. This
scheme  approximates  the Riemann shock propagation by assuming a region of
constant flow between the discontinuous surfaces. Our present  calculations
have  been  performed  using Relativistic HLLE (RHLLE) program to solve the
hydrodynamical equations \cite{rhlle1,rhlle2}. An important input for  this
method  is  the  equation  of  state  (EoS)  specifying  the  pressure as a
functional of the energy density and the number density  $p(\epsilon  ,n)$.
Once  an  EoS is specified, the hydrodynamical equations uniquely determine
the expansion dynamics. We obtain the  EoS  for  hadronic,  mixed  and  QGP
states  at  finite baryon density. In this study, the EoS for the QGP state
is taken from bag model consisting of u,d quarks and gluons. The  pressure,
energy and entropy densities are given by,

\begin{equation}
p=\frac{37\pi^2}{90} T^4+\frac{1}{9}(\mu T)^2+\frac{1}{162\pi^2}\mu^4-B
\end{equation}
\begin{equation}
p=(\epsilon-4B)/3~;~s=\frac{\partial P}{\partial T}|_\mu~;~ \epsilon=Ts+\mu n-p
\end{equation}

\noindent
In above, the $\mu$ is the baryo chemical potential representing the finite
baryon  density  effects. The hadronic EoS is obtained from a general class
of thermodynamically self consistent equations of the  interacting  nuclear
matter  that  reproduce the ground state matter properties. The present EoS
includes the $\sigma,\omega$ mesons  interacting  with  nucleons  and  anti
nucleons  (i.e.  masses  upto  1.0  GeV)  and a free gas of $\pi,\rho,\eta$
mesons. The hadron pressure ($p_{h}$), net baryon number density ($n$), the
scalar density ($\rho$) and the partial  pressures  of  baryon  and  mesons
($p_N,p_i$) are as in \cite{eos1} given by,

\begin{equation}
 p_h(T,\mu)=p_N+\Sigma_ip_i+n{\bf \nu}(n)
 -\int_0^n{\bf \nu}(n')dn'-\rho_s {\bf S}(\rho_s)
 +\int_0^{\rho_s}{\bf S}(\rho'_s)d{\rho'}_s
\end{equation}
\begin{equation}
 n(T,\mu)=\frac{g_N}{(2\pi^3)}\int d^3k
\left [f_T(E^*_k,\nu)-f_T(E^*_k,-\nu)\right ],
 ~~~\mbox{with}~~~~\nu=\mu-{\bf\nu}(n)
\end{equation}
\begin{equation}
\rho(T,\mu)=\frac{g_N}{(2\pi^3)}\int d^3k \frac{M^*}{E^*_k}
 \left[f_T(E^*_k,\nu)+f_T(E^*_k,-\nu)\right]
\end{equation}
\begin{equation}
 E^*_k=\sqrt{k^2+{M^*}^2} ~;~~
 M^*=M-{\bf S}(\rho_s) ~~;~~\\
 {\bf\nu}(n)=C_v^2n-C_d^2\rho_s ~;~ {\bf S}(\rho_s)=C_s^2\rho_s
\end{equation}
\begin{equation}
 p_N= T\frac{g_N}{(2\pi^3)}\int d^3k \left[\mbox{ln}( 1+e^{-(E^*_k-\nu)/T})+\mbox{ln}(1+e^{-(E^*_k+\nu)/T}\right] \nonumber\\
\end{equation}
\begin{equation}
 p_i(T;m_i)=-T\frac{g_i}{(2\pi^3)}\int d^3q \left[\mbox{ln}( 1-e^{-(\sqrt{q^2+m_i^2}/T)})\right]
\end{equation}

\noindent
In  the  above,  $f_T$ is the fermi distribution at temperature T and baryo
chemical potential $\mu$ representing the baryon  density.  $m_i$  are  the
meson  masses,  M  is the nucleon bare mass and $g_i,g_N$ are the meson and
nucleon spin-isospin degeneracies.  The  energy  density  ($\epsilon$)  and
entropy  ($s$)  are obtained using thermodynamic relations as in Eq. 6. The
effective chemical potential ${\bf \nu}(n)$ and the  effective  mass  $M^*$
are   obtained   using  the  parameters  given  in  $C_v^2=238.08GeV^{-2}$,
$C_s^2=296.05GeV^{-2}$, $C_d^2=0.183$  \cite{eos1}.  The  Eqs.(8,9,10)  are
solved  self  consistently  . The phase boundary in $T,\mu$ plane, shown in
Fig. 1(a), is obtained from the Gibbs conditions given by,

\begin{equation}
p_{h}=p_{q} ,~~~~~T_{h}=T_{q} ,~~~~~\mu_{h}=\mu_{q}
\end{equation}
The mixed phase for a given ($\epsilon$,n) is obtained from,
\begin{equation}
\epsilon=\lambda_q\epsilon_q(T^*)+(1-\lambda_q)\epsilon_h(T^*)~~~~~~\mbox{and}~~~~~~
n=\lambda_qn_q(T^*)+(1-\lambda_q)n_h(T^*)
\end{equation}

\noindent
In  the  above,  the  subscripts  $\mu,h$  denote  the  qgp,hadron  phases.
$\lambda_q$ is the fraction of volume occupied by  the  QGP  in  the  mixed
phase. The EoS specifies the pressure as a functional of p($\epsilon$,n) in
the  required region in ($\epsilon$,n) plane as shown in the Fig. 1 (nearly
same as Fig. 1 of \cite{eos1}). Fig. 1(a) shows the transition  temperature
and  chemical potential given by Eq. 13. The transition temperature at zero
baryon density is 169.9 MeV. Fig. 1(b) shows the energy  density  variation
and  the  Fig.  1(c)  shows the number density variation along the curve of
Fig. 1(a). Fig.1(d) shows the energy density and number  density  for  both
quark  and  hadron  phases for $T,\mu$ values along the curve of Fig. 1(a).
The hadron phase is the narrow strip in  the  semi-infinite  ($\epsilon$,n)
plane as shown in Fig. 1(d) by the mixed phase above and below by the $T=0$
line  of  hadron phase. However, owing to small life times of QGP and mixed
phases (10-20 fm/c), these phases occupy very small region  of  space  time
evolution diagrams which will be discussed later (Fig. 2). The $T^*$ of Eq.
14  are  determined  using Figs. 1(b,c,a). The constants used in Fig. 1 are
$\epsilon_0=0.1477\hbar^3 GeV^4$ and $n_0=0.16\hbar^3 GeV^3$.

\noindent
We  studied  the expansion dynamics at finite baryon density for two cases;
case (i) for SPS energy and case (ii) for the RHIC energy. The  case  (iii)
for  SPS energy at zero baryon density is studied to compare the effects of
finite baryon density. The initial energy density and the number density at
time $\tau=\tau_i$ have been assumed to be  of  wounded  nucleon  form,  as
given  by  $f(r)=\frac{3}{2}\sqrt{1-\frac{r^2}{R^2}}$  and R is the nuclear
radius taken to be 6.0 fm. The formation times are taken as $\tau_i$=1 fm/c
for SPS energy and $\tau_i$=0.6 fm/c for the  RHIC  energies  respectively.
The parameters for RHIC energies were taken from \cite{rhlle1}. The entropy
(s),  energy  and number densities corresponding to SPS case of 158 AGeV of
Pb+Pb collision is obtained by solving the simultaneous equations given by,

$$\pi R^2n_b\tau_i=\frac{dN_B}{dy}$$
$$  \pi R^2s\tau_i=3.6\frac{dN}{dy}$$

\noindent
We  have  taken  $\frac{dN_B}{dy}=80$,  and  $\frac{dN}{dy}=700$  from  the
experimental data, which corresponds to a baryon density of $n_b/n_0$=4.5 .
These space-time evolution of the plasma for the cases  of  SPS  energy  at
finite  baryon density is shown in Fig. 2 for different phases as indicated
in the figure in various colors.  The evolution of zero baryon  density  is
also  shown in Fig. 2, by thick colored  lines representing the boundary of
corresponding phases. The radial extension and the proper time (at z=0) are
in units of nuclear radius. The outermost envelope of the hadron  phase  in
Fig.  2  corresponds  to the freeze out isotherm taken to be 120.0 MeV. The
QGP phase region is almost the same for the two  cases  (slightly  less  at
finite  density).  It  can  be seen in Fig. 2 that at SPS energy the baryon
density decreases the life times of QGP and mixed phases of the  plasma  as
well  as  the freeze out time. The transverse extension is also affected by
the baryon density. Similar cases of space time evolution have been studied
for RHIC energy. At RHIC energies, the life times are  considerably  larger
together with the large transverse extension (see Fig.2 of \cite{rhlle1}).

\noindent
The  temperature,  chemical  potential,  constant  energy  density  and the
transverse velocity profiles in the space time, as the plasma expands, have
been studied for these cases. The initial temperatures are  larger  in  the
QGP  phase  which  in  time  go  over  to the constant temperature profiles
corresponding to mixed phases. Subsequently,  the  temperature  falls  very
rapidly  as  the plasma cools. The maximum temperatures (at r=0 for wounded
nucleon form) reached are approximately 239.06  MeV,  244.25  MeV  for  SPS
energies  for the cases of with and without baryon density respectively and
348.4 MeV at RHIC energy. The average energy densities  correspond  to  the
temperatures  of 214 MeV, 219.5 MeV at SPS energy for the cases of with and
without baryon density. These results are comparable to those of Table 1 of
\cite{huovinen}.\\

\noindent
The  energy  flow  in  space-time while in expansion is demonstrated by the
constant energy contours. These  contours  for  the  SPS  cases  have  been
compared in Fig. 3. The solid curves are with baryon density and the dashed
curves  are  for  zero  density.  The constant energy density values are as
mentioned in the figure in units if GeV/fm$^3$. It can  be  seen  from  the
figures that the energy flow gets slower in the presence of baryon density.
Similarly,  the constant transverse velocity profiles for various values of
$\beta=v/c$ have been shown for these cases in Fig. 4.

\par

\noindent
In  the following we discuss the static rates for photon emission needed to
calculate the total photon yields. The annihilation processes  $(q{\bar  q}
\rightarrow  g\gamma)$  and  the  QCD  Compton  $(qg \rightarrow q \gamma$,
${\bar q}g \rightarrow {\bar q} \gamma)$ processes have been considered for
the   thermal   photon   production   at   effective   one    loop    level
\cite{kapusta,trax1}.  The  photon  emission rates from quark matter at one
loop have been taken from the Traxler {\it et al.} \cite{trax1}, and at two
loop level for non zero chemical potentials as reported  in  \cite{dutta1}.
We  reported  the  photon production rate upto two loop level for a general
case  of  chemically  unsaturated  plasma  at  finite  baryon  density   in
\cite{dutta1}.  The  bremsstrahlung  and annihilation with scattering ($\bf
aws$) processes at two loop level have been considered.  At  finite  baryon
density,  the  photon  production  at  one  loop  level is dominated by the
compton process compared to annihilation process. At two  loop  level,  the
bremsstrahlung  radiation  is  affected  whereas  the  $\bf aws$ process is
insensitive to the baryon density presence.  The  bremsstrahlung  radiation
from  quark  is  enhanced and from anti quark is suppressed at finite baryo
chemical potential.

\par

\noindent
The   photon  emission  rates  from  the  hadron  phase  is  determined  by
considering   various   meson   reactions,   such    as    $\rho\rightarrow
\pi\pi\gamma$, $\pi\rho\rightarrow\pi\gamma$, $\pi\pi\rightarrow\rho\gamma,
\omega\rightarrow\pi\gamma$.   The  $\pi\rho\rightarrow\pi\gamma$  estimate
also includes the $a_1$ intermediate resonance. The numerical results of  a
detailed study of these decays \cite{kapusta} has been fitted by analytical
expressions  as a function of temperature and photon energy, as reported in
\cite{nadeau,steffen,peitz}. In the present work, we used these  analytical
formulae  for  the  photon  rate  calculations. It has been shown that from
temperature in the range of 100-200 MeV,  these  formulae  reproduce  quite
well  the  numerical  results  from  detailed  calculations  for the photon
energies from 200 MeV and upto 5.0 GeV. The one  dimensional  expansion  of
the  plasma  with  boost  invariance  has  been  already studied with these
emission rates with and without baryon  density.  It  was  found  that  the
finite   baryon   density   has  no  effect  on  the  total  photon  yields
\cite{dutta2}. The total photon yield integrated over the plasma space-time
hyper volume at zero rapidity is given by,

\begin{equation}
\frac{dN}{d^2P_Tdy}=\int^{\tau_f}_{\tau_i}\tau d\tau\int_0^{R_f}rdr\int_0^{2\pi}d\phi\int_{-\infty}^{\infty}d\eta
\left(\frac{EdR}{d^3p}|_{q}+\left(\lambda\frac{EdR}{d^3p}|_{q}+(1-\lambda)\frac{EdR}{d^3p}|_{h}\right)+\frac{EdR}{d^3p}|_{h}\right)
\end{equation}

\noindent
In  above, the subscripts $q$ and $h$ denote the photon emission rates from
QGP and hadron phases.  The  $\tau_f$  and  R$_f$  denote  the  freeze  out
(outermost)  boundary  shown in Fig. 2. The rates correspond to the photons
emitted in a fluid rest frame. In order to compare  with  the  experimental
photon  yields,  the  transverse velocity distributions of Fig. 4 have been
used to transform the photon energy as  given  by  $E_\gamma=p^\mu  u_\mu$.
Fig.  5 shows the photon transverse momentum rapidity density distributions
at zero rapidity for the 158 A GeV SPS energy for Pb+Pb collision  compared
with  WA98  data. The finite and zero density cases are shown in figure. As
seen in figure, the finite baryon density has very  little  effect  on  the
photon  spectrum,  though  the  yield  seems to be reduced marginally. This
result is surprising considering that the expansion dynamics is affected as
shown in Fig. 2. However, at these SPS energies, the  overall  plasma  life
time  is small and the expansion effects and modified emission rates due to
finite  chemical  potential  cancel  each  other.  However,  the   fit   to
experimental   data  is  slightly  improved  when  the  finite  density  is
considered (see Fig. 5). The prompt contribution in the present  study  has
been taken from \cite{wwang}. It has been noticed that the prompt estimates
of  \cite{gall}  and  \cite{wwang} differ by about a factor of two together
with roughly same slope in the region  of  1.5-4  GeV.  Therefore,  if  one
considers  the prompt contribution twice as large as given in \cite{wwang},
our  calculated  spectrum  at  finite  density  matches   well   with   the
experimental  data.  At  RHIC  energies,  the  baryon  number densities are
smaller and the finite density effects can be completely ignored. All these
studies have been repeated for  SPS  case  for  different  formation  times
$\tau_i$=1.1  fm/c,1.2 fm/c and  the  results  are  very  much similar. The
results for photon yields have not  changed  significantly  with  different
formation   times  studied.  The  present  study  was  partly  reported  in
\cite{svs1}.

\par

\noindent
The  photon  rates  from  QGP  phase at two loop level of \cite{dutta1} are
identical to the results of Aurenche ~~{\it et al.} \cite {auren1} for  the
equilibrated  plasma,  corrected  by  a factor of four. These rates and the
total photon transverse momentum distributions shown in  Fig.  5  for  zero
baryon  density  donot  include  the  LPM  effects.  The effective two loop
contributions to photon  production  rate  from  QGP  phase  consisting  of
bremsstrahlung  and  $\bf  aws$  processes  with and without LPM effect are
discussed in \cite{arnold1,arnold2}, especially as shown in Fig. 7 of  Ref.
\cite{arnold2}.  We have studied the consequences of photon rate inhibition
due to LPM effect on the space time integrated photon yields for  the  case
of zero density. For this purpose, we used the phenomenological expressions
of \cite{arnold2} that include LPM effects. The ratio of total photon yield
arising  from  the  two  loop  processes  with  and without LPM effect been
estimated and shown in Fig. 6. It has been found that the  LPM  effect  has
not  changed  the  photon transverse momentum distribution, due to the fact
that at SPS energy the quark matter contribution to photon  yield  is  very
less.  The LPM effect becomes important at higher energies as the {\bf aws}
contribution to photon emission is more suppressed. The total quark  matter
contribution  from  one  and  two  loop  processes without  LPM effects are
calculated and the relative percentage contribution of one  loop  processes
and the two loop processes have been obtained for the space-time integrated
yields.  It should be noted that the temperature and the chemical potential
dependence has been integrated out through the space time  integrations  of
Eq.  15 over the plasma volume. These  are also shown in Fig. 6 where short
dashes represent  oneloop/total  quark  matter  emission  and  long  dashes
similarly  for  the two loop contribution. The different color  curves show
the results for RHIC energy. At RHIC energies the LPM effects are important
as  the  QM  contribution  is  important  and  therefore  the  affects  the
calculated yields.

\vspace {1.cm}
\par
\noindent
{\bf Conclusion}

\noindent
The effects of finite baryon density and the (3+1) dimensional expansion of
the  quark  gluon  plasma  has  been studied for SPS and RHIC energies. The
transverse   expansion   has   been   obtained    following    relativistic
hydrodynamical  approach  and the relativistic HLLE method. The equation of
state for pressure functional has been obtained considering various  mesons
and  interacting  nucleons. The space time evolution of the QGP is shown to
be affected by the baryon density. The radial and the temporal extension of
the plasma is shorter in the presence of finite baryon density.  The  total
plasma  space  time integrated photon yields have been compared for various
cases in the presence of baryon density. Calculations at SPS energies  with
and  without  baryon density showed that the total photon yield is not very
sensitive to the baryon density, however individual contributions to photon
yields from various phases differ. The LPM effect is also found to  be  not
affecting  the photon yields for SPS energies, where as at RHIC energy this
effect is important.

\vspace {1.cm}
\par
\noindent
We acknowledge the fruitful discussions with Dr. S. Kailas.

\begin{figure}[!ht]
\begin{center}{
\hspace{-1.cm}
\begin{minipage}{18.cm}
\psfig{figure=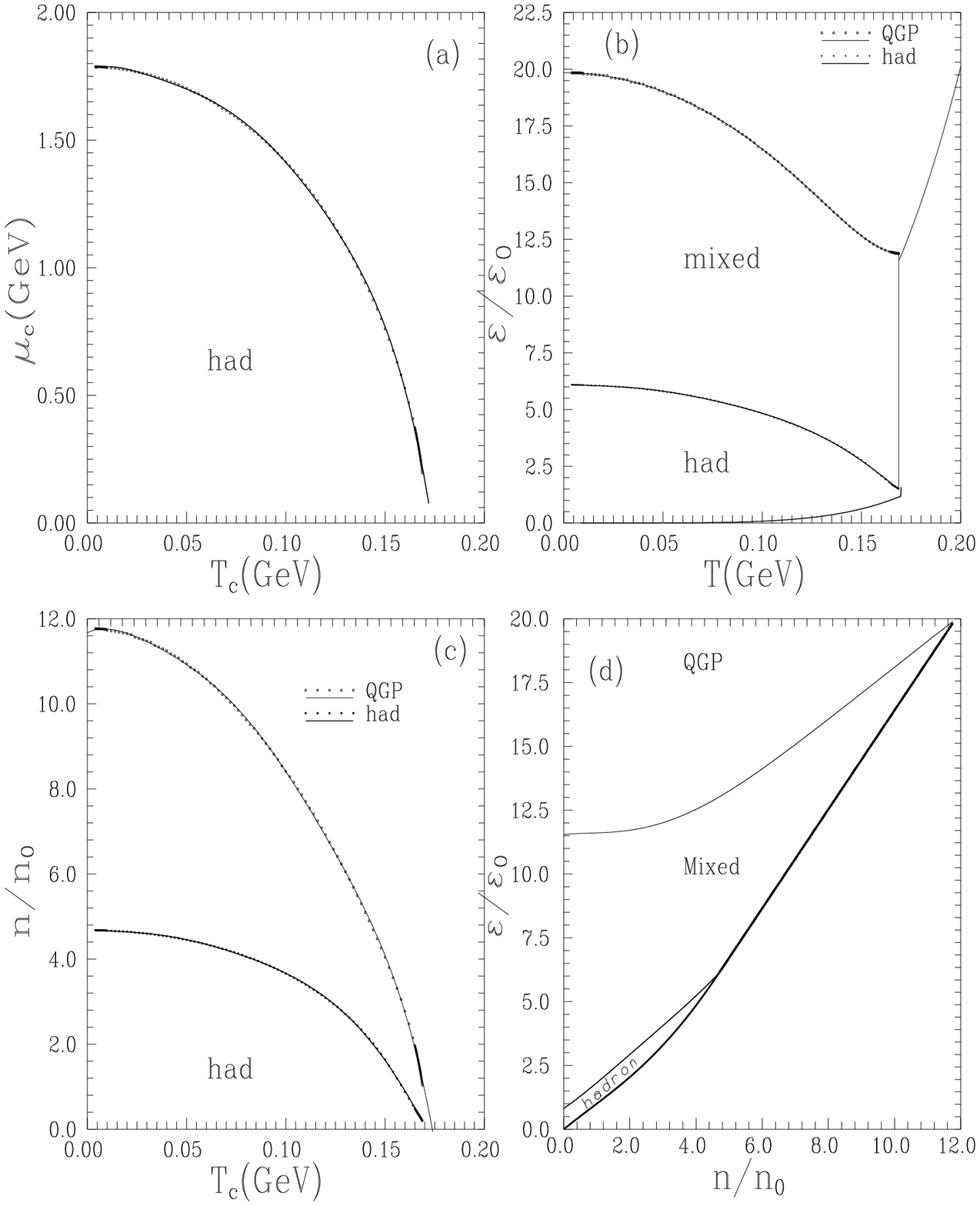,height=18.cm,width=16cm}
\end{minipage}
\caption{The phase diagram for QGP, Mixed and hadron
phases. The phase boundary is obtained from Gibbs conditions.}
}\end{center}
\vspace {-0.25cm}
\end{figure}
\begin{figure}[!ht]
\hspace{-1.cm}
\hspace{-1.cm}
\begin{minipage}{20.cm}
\psfig{figure=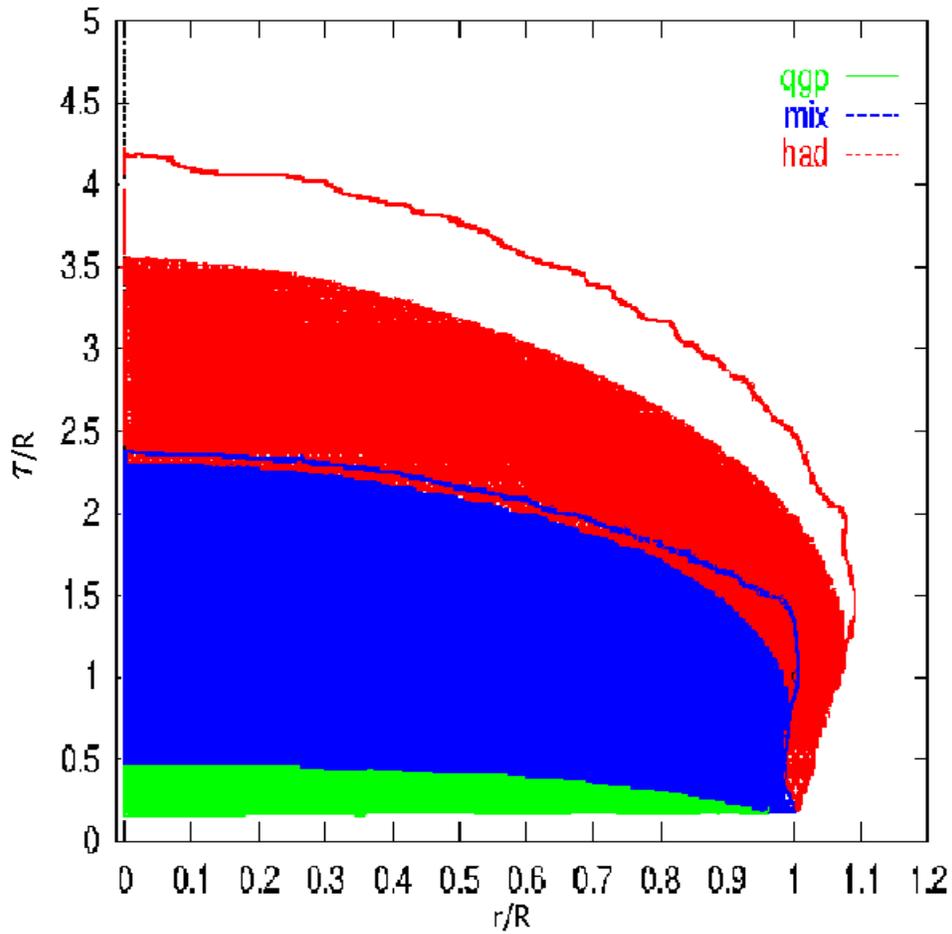,height=20.cm,width=16cm}
\vspace{-5.0cm}
\end{minipage}
\caption{ The space-time expansion of the QGP for SPS case. The
figure is for finite baryon density, and the corresponding colored
solid curves represent the extent of that phase for zero baryon density case.}

\end{figure}
\begin{figure}[!ht]
\begin{center}{
\hspace{-1.cm}
\begin{minipage}{18.0cm}
\psfig{figure=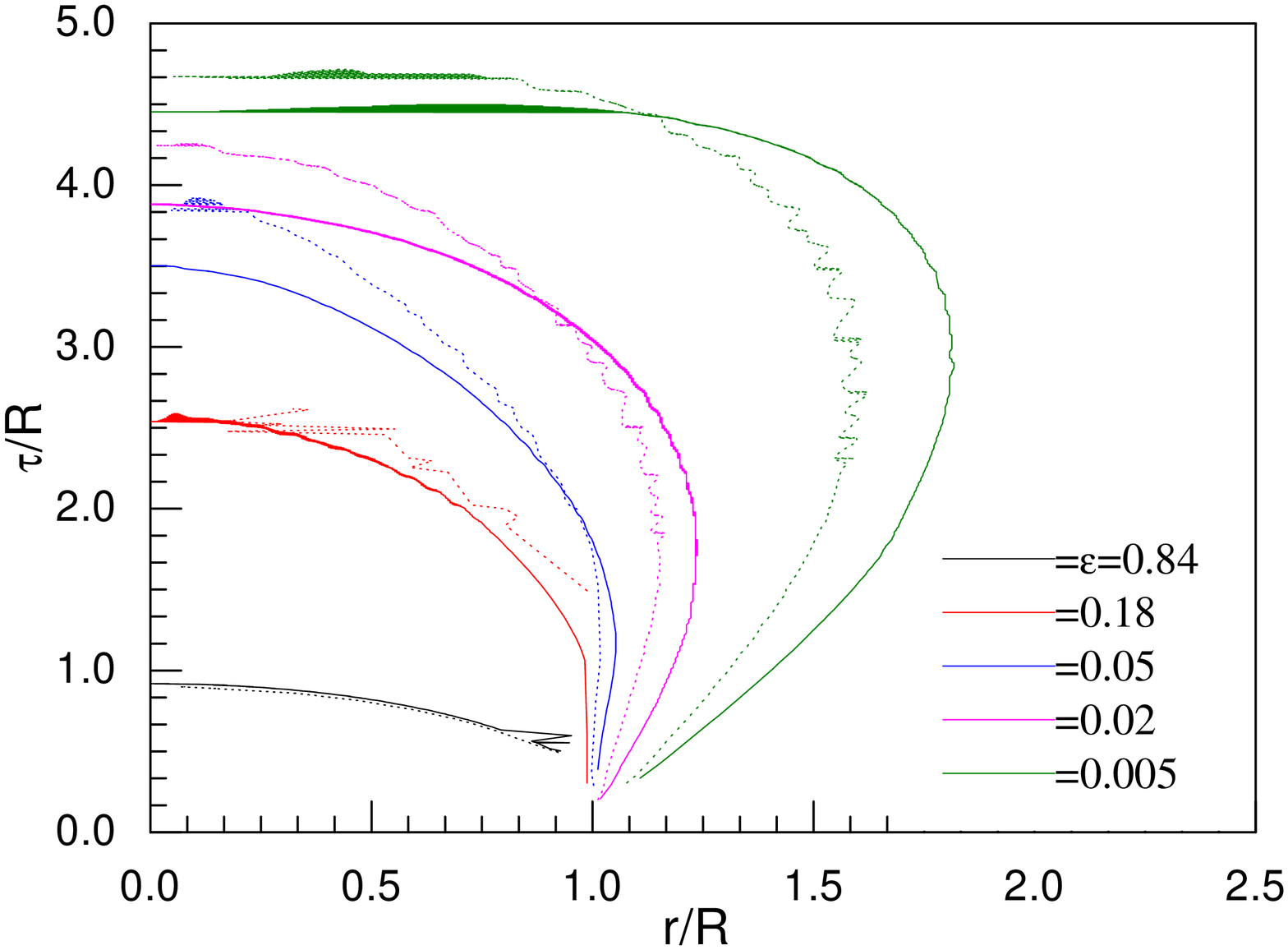,height=18.cm,width=16cm}
\end{minipage}
\caption{constant energy contours for the SPS energy. The solid curves
are with density effects, and the dashed curves are for zero density case.}
}\end{center}
\hspace {-0.25cm}
\end{figure}
\newpage
\begin{figure}[!ht]
\begin{center}{
\hspace{-1.cm}
\begin{minipage}{18.cm}
\psfig{figure=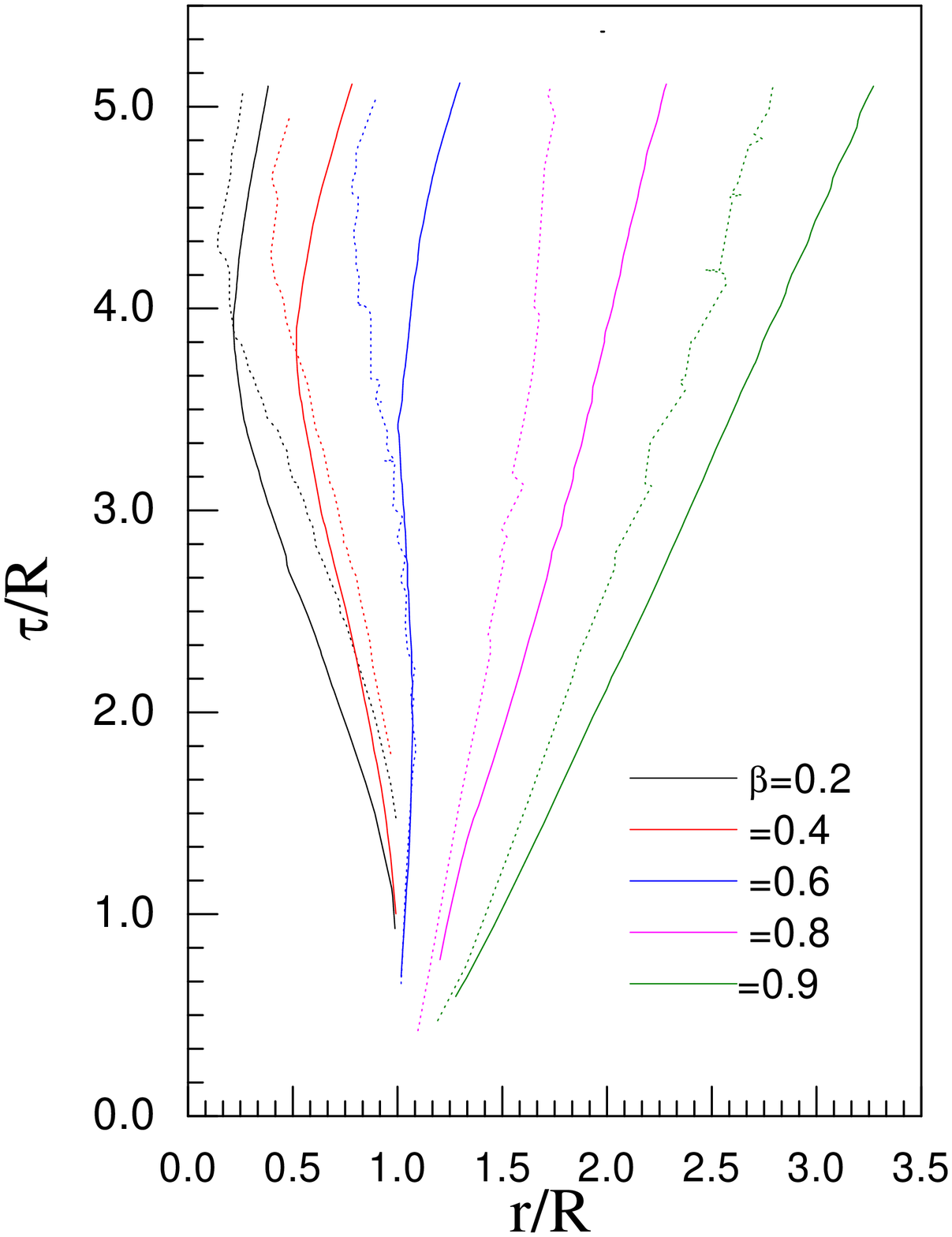,height=18.cm,width=16cm}
\end{minipage}
\caption{constant velocity contours for the SPS energy. The solid curves
are with density effects, and the dashed curves are for zero density case.}
}\end{center}
\hspace {-0.25cm}
\end{figure}
\begin{figure}[!ht]
\begin{center}{
\hspace{-1.cm}
\begin{minipage}{20.cm}
\psfig{figure=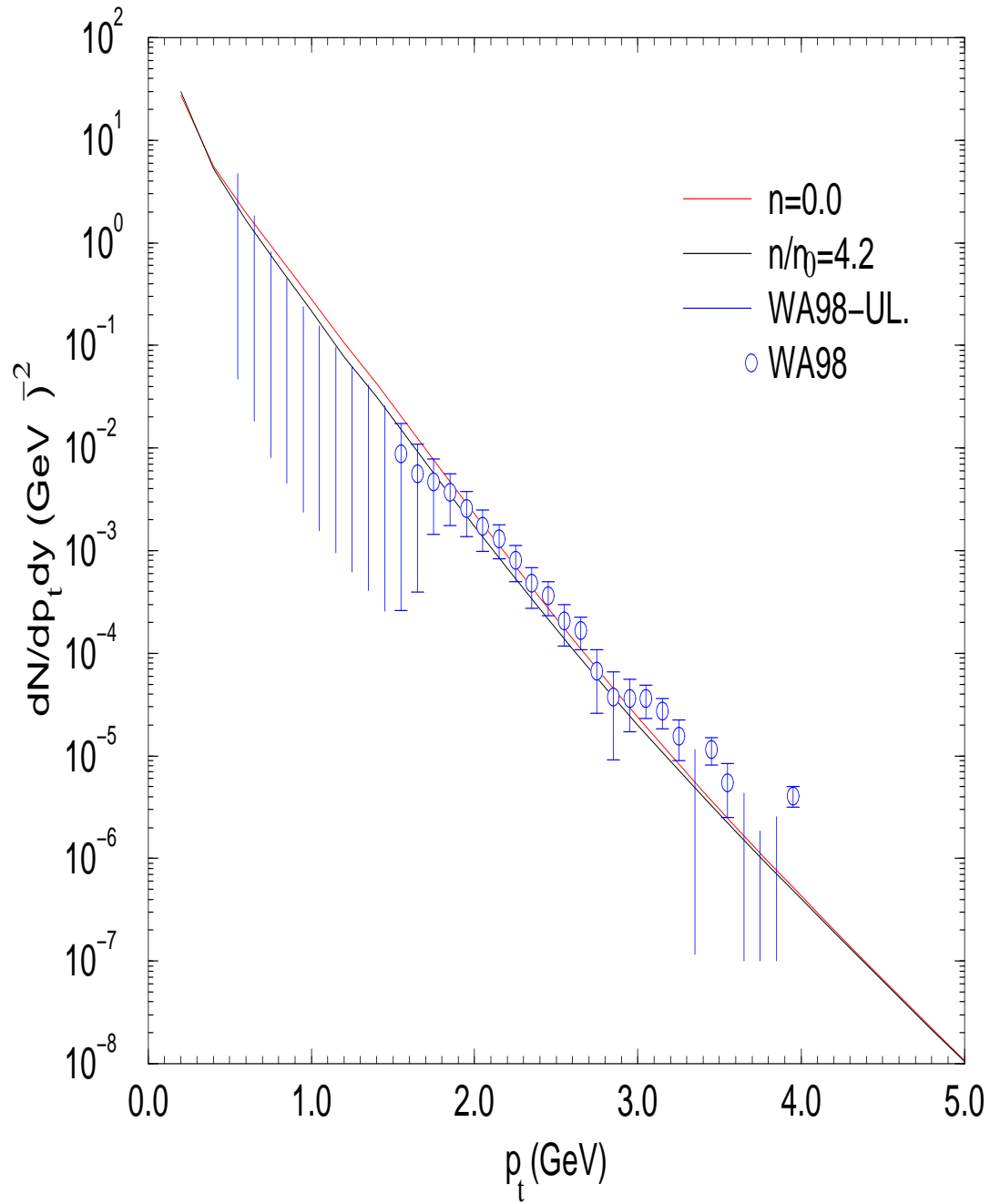,height=20.cm,width=16cm}
\end{minipage}
\caption{The total space time integrated photon transverse momentum distributions
for the SPS case with and without baryon density.}
}\end{center}
\hspace {-0.25cm}
\end{figure}
\begin{figure}[!ht]
\begin{center}{
\hspace{-1.cm}x
\begin{minipage}{18.cm}
\psfig{figure=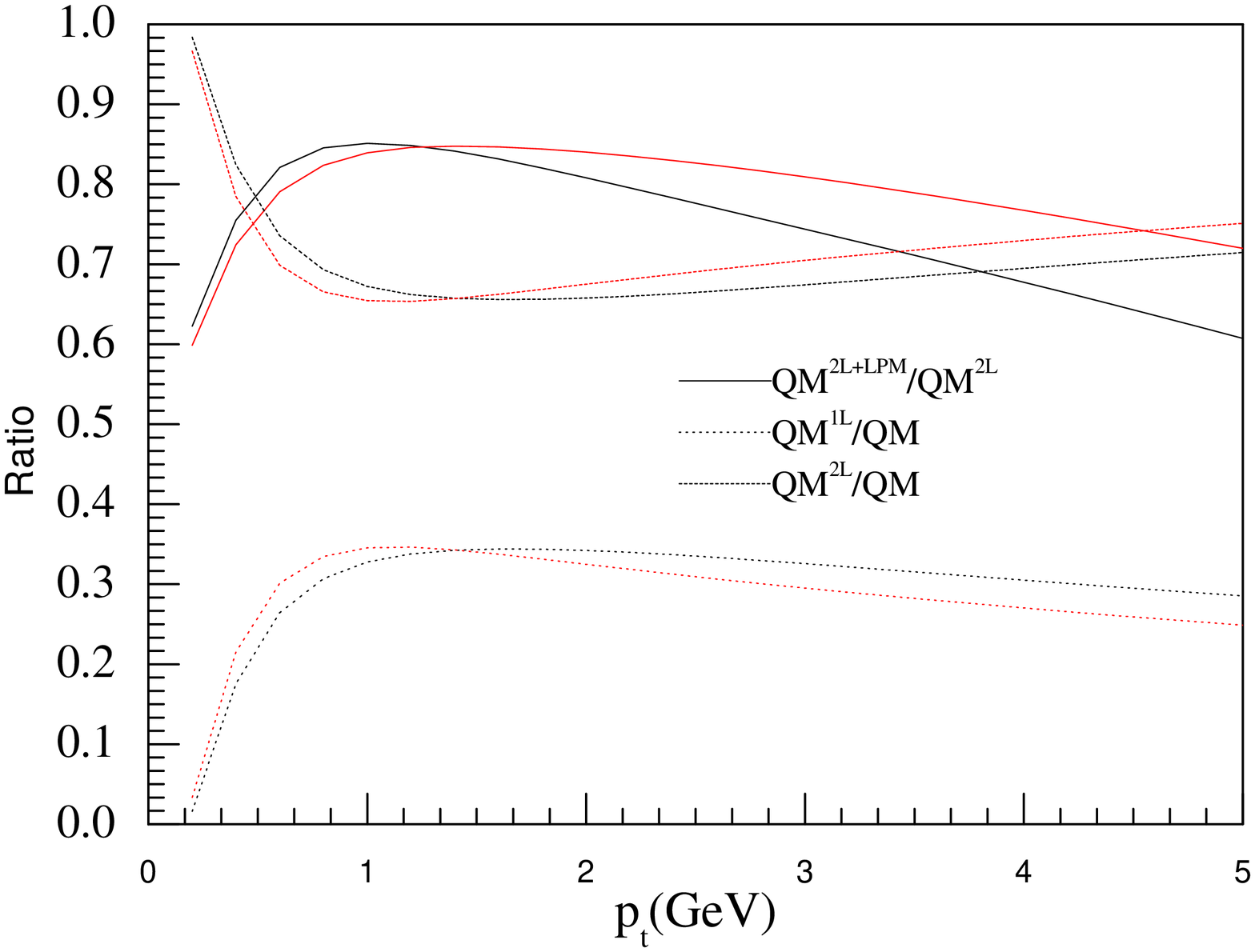,height=18.cm,width=16cm}
\end{minipage}
\caption{The suppression factor, for the two loop processes consisting
bremsstrahlung and aws processes from QGP phase, arising from LPM effects.
The LPM effect is shown by solid curves, the dotted curve  shows the relative
one loop contributions in total quark matter for the  photon yields and similarly  the
dashes  are for the relative two loop contribution.}
}\end{center}
\hspace {-0.25cm}
\end{figure}
\end{document}